\documentclass{article}

\usepackage[preprint, nonatbib]{nips_2018}

\usepackage[utf8]{inputenc} % allow utf-8 input
\usepackage[T1]{fontenc}    % use 8-bit T1 fonts
\usepackage{hyperref}       % hyperlinks
\usepackage{url}            % simple URL typesetting
\usepackage{booktabs}       % professional-quality tables
\usepackage{amsfonts}       % blackboard math symbols
\usepackage{nicefrac}       % compact symbols for 1/2, etc.
\usepackage{microtype}      % microtypography
\usepackage{xcolor}
\usepackage{bm}
\usepackage{amsmath}
\usepackage{graphicx}

\usepackage{amsmath}
\usepackage{float}
\usepackage[export]{adjustbox}
\usepackage{subcaption}

\usepackage{makecell}

\usepackage{todonotes}
\title{Deep Reinforcement Learning for Trading}

\author{
  Zihao~Zhang, Stefan~Zohren, and Stephen~Roberts\\
  Department of Engineering Science, \\
  Oxford-Man Institute of Quantitative Finance,
  University of Oxford
}

\begin{document}

\maketitle

\begin{abstract}

We adopt Deep Reinforcement Learning algorithms to design trading strategies for continuous futures contracts. Both discrete and continuous action spaces are considered and volatility scaling is incorporated to create reward functions which scale trade positions based on market volatility. We test our algorithms on the 50 most liquid futures contracts from 2011 to 2019, and investigate how performance varies across different asset classes including commodities, equity indices, fixed income and FX markets. We compare our algorithms against classical time series momentum strategies, and show that our method outperforms such baseline models, delivering positive profits despite heavy transaction costs. The experiments show that the proposed algorithms can follow large market trends without changing positions and can also scale down, or hold, through consolidation periods.
\end{abstract}

%%%%%%%%%%%%%%%%%%%%%%%%%%

\section{Introduction}

Financial trading has been a widely researched topic and a variety of methods have been proposed to trade markets over the last few decades. These include fundamental analysis \cite{graham1934security}, technical analysis \cite{murphy1999technical} and algorithmic trading \cite{chan2009quantitative}. Indeed, many practitioners use a hybrid of these techniques to make trades \cite{schwager2017complete}. Algorithmic trading has arguably gained most recent interest and accounts for about 75\% of trading volume in the United States stock exchanges \cite{chan2009quantitative}. The advantages of algorithmic trading are widespread, ranging from strong computing foundations, faster execution and risk diversification. One key component of such trading systems is a predictive signal that can lead to alpha (excess return) and, to this end, mathematical and statistical methods are widely applied. However, due to the low signal-to-noise ratio of financial data and the dynamic nature of markets, the design of these methods is non-trivial and the effectiveness of commonly derived signals varies through time.

%Time series momentum has shown strong predictability in the literature \cite{moskowitz2012time, hurst2017century} with evidence of consistent positive returns generated across very different asset classes and time periods \cite{lim2019enhancing}. The work of \cite{moskowitz2012time} suggests that strong price trends persist and, in particular, the past 12-month return of an instrument is a positive predictor of its future return. Ground on this observation, a group of trading strategies has been proposed to profit from trending markets \cite{kim2016time, baltas2017demystifying, harvey2018impact}, and volatility scaling \cite{moskowitz2012time, harvey2018impact} is adopted to scale up trades positions while volatility is low, and vice versa . However, the performance of time series momentum strategies has deteriorated since 2015 as markets tend to move sideways, leading to more false breakouts and longer consolidation periods. 

In recent years, machine learning algorithms have gained much popularity in many areas, with notable successes in  diverse application domains including image classification \cite{krizhevsky2012imagenet} and natural language processing \cite{collobert2011natural}. Similar techniques have also been applied to financial markets in an attempt to find higher alpha (see \cite{zhang2018bdlob, zhang2019deeplob, zhang2019extending, tsantekidis2017forecasting, sirignano2019universal, zhang2019deeplob} for a few examples using such techniques in the context of high-frequency data). Most research focuses on regression and classification pipelines in which excess returns, or market movements, are predicted over some (fixed) horizons. However, there is little discussion related to transforming these predictive signals into actual trade positions (see \cite{lim2019enhancing} for an attempt to train a deep learning model to learn positions directly). Indeed such a mapping is non-trivial. As an example, predictive horizons are often relatively short (one day or a few days ahead if using daily data), however large trends can persist for weeks or months with some moderate consolidation periods. We therefore not only need a signal with good predictive power but also a signal that can consistently produce good directional calls.

In this work, we report on Reinforcement Learning (RL) \cite{sutton1998introduction} algorithms to tackle above problems. Instead of making predictions - followed by a trade decision based on predictions, we train our models to directly output trade positions, bypassing the explicit forecasting step. Modern portfolio theory \cite{arrow1971theory, pratt1978risk, ingersoll1987theory} implies that, given a finite time horizon, an 
investor chooses actions to maximise some expected utility of final wealth:
\begin{equation} \label{eq:wealth}
\mathbb{E}[U(W_T)] = \mathbb{E}\left[U\left(W_0 + \sum_{t=1}^T \delta W_t\right)\right],
\end{equation}
where $U$ is the utility function, $W_T$ is the final wealth over a finite horizon $T$ and $\delta W_t$ represents the change in wealth. The performance of the final wealth measure depends upon sequences of interdependent actions where optimal trading decisions do not just decide immediate trade returns but also affect subsequent future returns. As mentioned in \cite{merton1969lifetime, ritter2017machine}, this falls under the framework of optimal control theory \cite{kirk2012optimal} and forms a classical sequential decision making process. If the investor is risk-neutral, the utility function becomes linear and we only need to maximise the expected cumulative trades returns, $\mathbb{E}(\sum_{t=1}^T \delta W_t)$ and we  observe that the problem fits exactly with the framework of RL,  the goal of which is to maximise some expected cumulative rewards via an agent interacting with an uncertain environment. Under the RL framework, we can directly map different market situations to trade positions and conveniently include market frictions, such as commissions, to our reward functions, allowing trading performance to be  directly optimised. 

%\paragraph{Contributions}

In this work, we adopt state-of-art RL algorithms to the aforementioned problem setting, including Deep Q-learning Networks (DQN) \cite{mnih2013playing, van2016deep}, Policy Gradients (PG) \cite{williams1992simple} and Advantage Actor-Critic (A2C) \cite{mnih2016asynchronous}. Both discrete and continuous action spaces are explored in our work, and we improve reward functions with volatility scaling \cite{moskowitz2012time, harvey2018impact} to scale up trade positions while volatility is low, and vice versa. We show the robustness and consistency of our algorithms by testing on the 50 most liquid futures contracts \cite{pinnacle} between 2011 and 2019. Our dataset consists of different asset classes including commodities, equity indexes, fixed income and FX markets. We compare our method with classical time series momentum strategies, and our method outperforms baselines models and generates positive returns in all sectors despite heavy transaction costs. Time series strategies work well in trending markets, like fixed income markets, however suffer losses in FX markets where directional moves are less usual. Our experiments show that our algorithms can monetise on large moves without changing positions and also deal with markets that are more ``mean-reverting''.

\paragraph{Outlines}
The remainder of the paper is structured as follows. Section~\ref{literature} introduces the current literature and Section~\ref{method} presents our methodology. In Section~\ref{experiment}, we compare our method with baseline algorithms and show how trading performance varies among different asset classes. A description of dataset and training methodology is also included in this section. Section~\ref{conclusion} concludes our findings and discusses extensions of our work.

\section{Literature Review}
\label{literature}

In this section, we review some classical trading strategies and discuss how RL has been applied to this field. Fundamental analysis aims to measure the intrinsic value of a security by examining economic data so investors can compare the security's current price with estimates to see if the security is undervalued or overvalued. One of the established strategies is called CAN-SLIM \cite{singhvi1988make} and it is based on a major study of market winners from 1880 to 2009. However, a common criticism of fundamental analysis is that the timing of enter and exit of trades is not specified. Even the markets move towards the estimated price, a bad timing of entering the trades could lead to huge drawdowns and  such moves in account values are often not bearable to investors, shaking them out of the markets. Technical analysis is in contrast to fundamental analysis where a security's historical price data is used to study price patterns. Technicians place trades based on a combination of indicators such as the Relative Strength Index (RSI) and Bollinger Bands. However, due to the lack of analysis on economic or market conditions, the predictability of these signals is not strong, often leading to false breakouts. 

Algorithmic trading is a more systematic approach that involves mathematical modelling and automated execution. Examples include trend-following \cite{schwager2017complete}, mean-reversion \cite{chan2013algorithmic}, statistical arbitrage \cite{chan2009quantitative} and delta-neutral trading strategies \cite{michaud1989markowitz}. We mainly review time series momentum strategies by \cite{moskowitz2012time} as we benchmark our models against their algorithms. Their work developed a very robust trading strategy by simply taking the sign of returns over the last year as a signal and demonstrated profitability for every contract considered within 58 liquid instruments over 25 years. Thenceforth, a number of methods \cite{baltas2017demystifying, baz2015dissecting, rohrbach2017momentum, lim2019enhancing} have been proposed to enhance this strategy by estimating the trend and map them to actual trade positions. However, these strategies are designed to profit from large directional moves but can suffer huge losses if markets move sideways as the predictability of these signals deteriorates and excess turnover erodes profitability. In our work, we adopt time series momentum features along with technical indicators to represent state space and obtain trade positions directly using RL. The idea of our representation is simple: extracting information across different momentum features and outputting positions based on the aggregated information.

%However, the core of these strategies is to profit from large infrequent directional moves so we always stay in the markets in case a large trend is missed. As a result, the winning rate of these trend following systems is usually low about 33\% \cite{tharp2007trade}, but the large profits made is enough to cover many small losses. However, if markets move sideways, these strategies suffer huge losses as the predictivity of these signals deteriorates and excess turnovers erodes our profitability. In our work, we optimise trades positions by using different time series momentums. Our experiments show that our model can scale up to large directional moves and also scale down during consolidation periods.   

The current literature on RL in trading can be categorized into three main methods: critic-only, actor-only and actor-critic approach \cite{fischer2018reinforcement}. The critic-approach, mainly DQN, is the most published method in this field \cite{bertoluzzo2012testing, jin2016portfolio, tan2011stock, huang2018financial, ritter2017machine} where a state-action value function, $Q$, is constructed to represent how good a particular action is in a state. Discrete action spaces are adopted in these works and an agent is trained to fully go long or short a position. However, a fully invested position is risky during high volatility periods, exposing one to severe risk when opposite moves occur. Ideally, one would like to scale up or down positions according to current market conditions. Doing this requires one to have large action spaces, however, the critic-approach suffers from large action spaces as we need to assign a score for each possible action.

The second most common approach is the actor-only approach \cite{moody1998performance, moody2001learning, lim2019enhancing, deng2016deep} where the agent directly optimises the objective function without computing the expected outcomes of each action in a state. Because a policy is directly learned, actor-only approaches can be generalised to continuous action spaces. In the work of \cite{moody2001learning, lim2019enhancing}, offline batch gradient ascent methods can be used to optimise the objective function, like profits or Sharpe ratio, as it is differentiable end to end. However, this is different from standard RL actor-only approaches where a distribution needs to be learned for the policy. In order to study the distribution of a policy, the Policy Gradient Theorem \cite{sutton1998introduction} and Monte Carlo methods \cite{metropolis1949monte} are adopted in the training and models are updated until the end of each episode. We often experience slow learning and need a lot of samples to obtain an optimal policy as individual bad actions will be considered ``good'' as long as the total rewards are good, taking long times to adjust these actions. 

The actor-critic approach forms the third category and aims to solve the above learning problems by updating the policy in real time. The key idea is to alternatively update two models where one, actor, controls how an agent performs given the current state, and the other, critic, measures how good the chosen action is. However, this approach is the least studied method in financial applications with limited works \cite{li2007short, bekiros2010heterogeneous, xiong2018practical}. We aim to supplement the literature and study the effectiveness of this approach in trading. For a more detailed discussion on state, action spaces and reward functions, interested readers are pointed to the survey \cite{fischer2018reinforcement}. Other important financial applications such as portfolio optimisation and trade execution are also included in this work. 

\section{Methodology}
\label{method}

In this section, we introduce our setups including state, action spaces and reward functions. We describe three RL algorithms used in our work, Deep Q-learning Networks (DQN), Policy Gradients (PG) and Advantage Actor-Critic (A2C) methods.

\subsection{Markov Decision Process Formalisation}

We can formulate the trading problem as a Markov Decision Process (MDP) where an agent interacts with the environment at discrete time steps. At each time step $t$, the agent receives some representation of the environment denoted as a state $S_t$. Given this state, an agent chooses an action $A_t$, and based on this action, a numerical reward $R_{t+1}$ is given to the agent at the next time step, and the agent finds itself in a new state $S_{t+1}$. The interaction between the agent and the environment produces a trajectory $\tau =[ S_0, A_0, R_1, S_1, A_1, R_2, S_2, A_2, R_3, \cdots]$. At any time step $t$, the goal of RL is to maximise the expected return (essentially the expected discounted cumulative rewards) denoted as $G_t$ at time $t$:
\begin{equation} \label{eq:rl_return}
    G_t = \sum_{k=t+1}^{T} \gamma^{k-t-1} R_k
\end{equation}
where $\gamma$ is the discounting factor. If the utility function in Equation~\ref{eq:wealth} has a linear form and we use $R_t$ to represent trade returns, we can see that optimising $\mathbb{E}(G)$ is equivalent to optimising our expected wealth. 

\paragraph{State Space}

In the literature, many different features have been used to represent state spaces. Among these features, a security's past price is always included and the related technical indicators are often used \cite{fischer2018reinforcement}. In our work, we take past price, returns ($r_t$) over different horizons and technical indicators including Moving Average Convergence Divergence (MACD) \cite{baz2015dissecting} and the Relative Strength Index (RSI) \cite{wilder1978new} to represent states. At a given time step, we take the past 60 observations of each feature to form a single state. A list of our features is below:
\begin{itemize}
\item Normalised close price series,
\item Returns over the past month, 2-month, 3-month and 1-year periods are used. Following \cite{lim2019enhancing}, we normalise them by daily volatility adjusted to a reasonable time scale. As an example, we normalise annual returns as $r_{t-252 ,t} / (\sigma_t \sqrt{252})$ where $\sigma_t$ is computed using an exponentially weighted moving standard deviation of $r_t$ with a 60-day span,
\item MACD indicators are proposed in \cite{baz2015dissecting} where:
\begin{equation} \label{eq:macd}
\begin{split}
& \text{MACD}_t = \frac{q_t}{\text{std}(q_{t-252:t})} \\
& q_t = (m(S) - m(L)) /  \text{std} (p_{t-63:t}) \\
\end{split}
\end{equation}
where std$(p_{t-63:t})$ is the 63-day rolling standard deviation of prices $p_t$ and $m(S)$ is the exponentially weighted moving average of prices with a time scale $S$,
\item The RSI is an oscillating indicator moving between 0 and 100. It indicates the oversold (a reading below 20) or overbought (above 80) conditions of an asset by measuring the magnitude of recent price changes. We include this indicator with a look back window of 30 days in our state representations. 
\end{itemize}

\paragraph{Action Space}

We study both discrete and continuous action spaces. For discrete action spaces, a simple action set of 
$\{-1, 0, 1\}$ is used and each value represents the position directly, i.e.\ $-1$ corresponds to a maximally short position, $0$ to no holdings and $1$ to a maximally long position. This representation of action space is also known as target orders \cite{huang2018financial} where a trade position is the output instead of the trading decision. Note that if the current action and next action are the same, no transaction cost will occur and we just maintain our positions. If we move from a fully long position to a short position, transaction cost will be doubled. The design of continuous action spaces is very similar to discrete case where we still output trade positions but allow actions to be any value between -1 and 1 ($A_t \in [-1,1]$).

\paragraph{Reward Function}

The design of the reward function depends on the utility function in Equation~\ref{eq:wealth}. In this work, we let the utility function be profits representing a risk-insensitive trader, and the reward $R_t$ at time $t$ is:
\begin{equation} \label{eq:add_profit}
R_t = \mu \Big [ A_{t-1} \frac{\sigma_{tgt}}{\sigma_{t-1}} r_t - \mathrm{bp} \ p_{t-1} \ \Big|   \frac{\sigma_{tgt}}{\sigma_{t-1}} A_{t-1} - \frac{\sigma_{tgt}}{\sigma_{t-2}} A_{t-2}\Big|  \Big ]
\end{equation}

where $\sigma_{tgt}$ is the volatility target and $\sigma_{t-1}$ is an ex-ante volatility estimate calculated using an exponentially weighted moving standard deviation with a 60-day window on $r_{t}$. This expression forms the volatility scaling in \cite{moskowitz2012time, harvey2018impact, lim2019enhancing} where our positions is scaled up when market volatility is low and scaled down vice versa. In addition, given a volatility target, our reward $R_t$ is mostly driven by our actions instead of being heavily affected by market volatility. We can also consider the volatility scaling as normalising rewards from different contracts to a same scale. Since our data consists of 50 futures contracts with different price ranges, we need to normalise different rewards to a same scale for training and also for portfolio construction. $\mu$ is a fixed number per contract at each trade and we set it to 1. Note that our transaction cost term also includes a price term $p_{t-1}$. This is necessary as again we work with many contracts and each contract has different costs, so we represent transaction cost as a fraction of traded value. The constant, bp (basis point) is the cost rate and 1bp=0.0001. As an example, if the cost rate is 1bp, we need to pay \$0.1 to buy one unit of a contract priced at \$1000.

We define $r_t = p_{t} - p_{t-1}$ and this expression represents additive profits. Additive profits are often used if a fixed number of shares or contracts is traded at each time. If we want to trade a fraction of our accumulated wealth at each time, multiplicative profits should be used and $r_t = p_{t} / p_{t-1} - 1$. In this case, we also need to change the expression of $R_t$ as $R_t$ represents the percentage of our wealth. The exact form can be found in \cite{moody1998performance}. We stick to additive profits in our work as logarithmic transformation needs to be taken for multiplicative profits to have cumulative rewards required by the RL setup, but logarithmic transformation penalises large wealth growths.

\subsection{RL Algorithms}

\textbf{Deep Q-learning Networks (DQN)} 
A DQN approximates the state-action value function (Q function) to estimates how good it is for the agent to perform a given action in a given state by adopting a neural network. Suppose our Q function is parameterised by some $\theta$. We minimise the mean squared error between the current and target Q to derive the optimal state-action value function:
\begin{equation}
\begin{split}
&L(\theta) = \mathbb{E} [(Q_{\theta} (S, A) -  Q^{'}_{\theta} (S, A))^2] \\
&Q^{'}_{\theta} (S_t, A_t) = r + \gamma \ \text{argmax}_{A'} Q_{\theta}(S_{t+1}, A_{t+1})
\end{split}
\end{equation}
%\szcom{there is an $S'$ on the right and $S$ on the left in the last equation: check!}
% , and $S'$ and $A'$ are state and action at the next step
where $L(\theta)$ is the objective function. A problem is that the training of a vanilla DQN is not stable and suffers from variability. Many improvements have been made to stabilise the training process, and we adopt the following three strategies, fixed Q-targets \cite{van2016deep}, Double DQN \cite{hasselt2010double} and Dueling DQN \cite{wang2016dueling} to improve the training in our work. Fixed Q-targets and Double DQN are used to reduce policy variances and to solve the problem of ``chasing tails'' by using a separate network to produce target values. Dueling DQNs separate the Q-value into state value and the advantage of each action. The benefit of doing this is that the value stream gets more updates and we learn a better representation of the state values. 

\textbf{Policy Gradients (PG)} 
The PG aims to maximise the expected cumulative rewards by optimising the policy directly. If we use a neural network with parameters $\theta$ to represent the policy, $\pi_{\theta}(A|S)$, we can generate a trajectory $\tau = [S_0, A_0, R_1, S_1, \cdots, S_t, A_t]$ from the environment and obtain a sequence of rewards. We maximise the expected cumulative rewards $J(\theta)$ by using gradient ascent to adjust $\theta$:
\begin{equation}
\begin{split}
J(\theta) &= \mathbb{E} [ \sum_{t=0}^{T-1} R_{t+1} | \pi_{\theta} ] \\
\nabla_{\theta} J(\theta) &=  \sum_{t=0}^{T-1} \nabla_{\theta} \log \pi_{\theta}(A_t | S_t) G_t
\end{split}
\end{equation}

where $G_t$ is defined in Equation~\ref{eq:rl_return}. Compared with DQN, PG learns a policy directly and can output a probability distribution over actions. This is useful if we want to design stochastic policies or work with continuous action spaces. However, the training of PG uses Monte Carlo methods to sample trajectories from the environment and the update is done only when the episode finishes. This often results in slow training and it can get stuck at a (sub-optimal) local maximum.

\textbf{Advantage Actor-Critic (A2C)} 
The A2C is proposed to solve the training problem of PG by updating the policy in real time. It consists of two models: one is an actor network that ouputs the policy and the other is a critic network that measures how good the chosen action is in the given state. We can update the policy network $\pi(A|S, \theta)$ by maximising the objective function:
\begin{equation}
J(\theta) = \mathbb{E}[ \log \pi(A|S, \theta) A_{adv}(S,A)]
\end{equation}

where $A_{adv}(S,A)$ is the advantage function defined as:
\begin{equation}
A_{adv}(S_t, A_t) = R_t + \gamma V(S_{t+1}| w) - V(S_t|w)
\end{equation}

In order to calculate advantages, we use another network, the critic network, with parameters $w$ to model the state value function $V(s|w)$, and we can update the critic network using gradient descent to minimise the TD-error:
\begin{equation}
J(w) = \left ( R_t + \gamma V(S_{t+1}|w) - V(S_t|w) \right )^2
\end{equation}

The A2C is most useful if we are interested in continuous action spaces as we recude the policy variance by using the advantage function and update the policy in real time. The training of A2C can be done synchronously or asynchronously (A3C). In this work, we adopt the synchronous approach and execute agents in parallel on multiple environments.

%\subsection{Model Architecture}
%%%%%%%%%%%%%%%%%%%%%%%%%%%%%%%%%%%%%%%%%%%%%%
%%%%%%%%%%%%%%%%%%%%%%%%%%%%%%%%%%%%%%%%%%%%%%
%%%%%%%%%%%%%%%%%%%%%%%%%%%%%%%%%%%%%%%%%%%%%%

\section{Experiments}
\label{experiment}

\subsection{Description of Dataset}

We use data on 50 ratio-adjusted continuous futures contracts from the Pinnacle Data Corp CLC Database \cite{pinnacle}. Our dataset ranges from 2005 to 2019, and consists of a variety of asset classes including commodity, equity index, fixed income and FX. A full breakdown of the dataset can be found in Appendix A. We retrain our model at every 5 years, using all data available up to that point to optimise the parameters. Model parameters are then fixed for the next 5 years to produce out-of-sample results. In total, our testing period is from 2011 to 2019.

\subsection{Baseline Algorithms}

We compare our methods to the following baseline models including classical time series momentum strategies:
\begin{itemize}
\item Long Only
\item Sign(R) \cite{moskowitz2012time, lim2019enhancing}:
\begin{equation}
A_t = \text{sign}(r_{t-252:t})
\end{equation}
\item MACD Signal \cite{baz2015dissecting}
\begin{equation}
\begin{split}
A_t &= \phi(\text{MACD}_t)\\
\phi( \text{MACD}) &= \frac{\text{MACD} \ \text{exp}(-\text{MACD}^2/4)}{0.89}
\end{split}
\end{equation}
where $\text{MACD}_t$ is defined in Equation~\ref{eq:macd}. We can also take multiple singals with different time-scales and average them to give a final indicator:
\begin{equation}
\tilde{\text{MACD}_t} = \sum_k \text{MACD}_t(S_k, L_k)
\end{equation}
where $S_k$ and $L_k$ define short and long time-scales, namely $S_k \in \{8,16,32\}$ and $L_k \in \{24,48,96 \}$ as in \cite{lim2019enhancing}.
\end{itemize}

We compare the above baseline models with our RL algorithms, DQN, PG and A2C. DQN and PG have discrete action spaces $\{-1, 0, 1\}$, and A2C has a continous action space $[-1, 1]$.

\subsection{Training Schemes for RL}

In our work, we use Long Short-term Memory (LSTM) \cite{hochreiter1997long} neural networks to model both actor and critic networks. LSTMs are traditionally used in natural language processing, but many recent works have applied them to financial time-series \cite{fischer2018deep, tsantekidis2017using, bao2017deep, di2016artificial}, in particular, the work of \cite{lim2019enhancing} shows that LSTMs deliver superior performance on modelling daily financial data. We use two-layer LSTM networks with 64 and 32 units in all models, and Leaky Rectifying Linear Units (Leaky-ReLU) \cite{maas2013rectifier} are used as activation functions. As our dataset consists of different asset classes, we train a separate model for each asset class. It is a common practice to train models by grouping contracts within the same asset class, and we find it also improves our performance. 

We list the value of hyperparameters for different RL algorithms in Table~\ref{table:parameter}. We denote the learning rates for critic and actor networks as $\alpha_{\mathrm{critic}}$ and $\alpha_{\mathrm{actor}}$. The Adam optimiser \cite{kingma2014adam} is used for training all networks, and batch size means the size of mini-batch used in gradient descent. As previously introduced, $\gamma$ is the discounting factor and bp is the cost rate used in training. We can treat bp as a regularising term as a large bp penalises turnovers and lets agents maintain current trade positions. The memory size shows the size of the buffer for experience replay, and we update the parameters of our target network in DQN at every $\tau$ steps.

\begin{table}[H]
\caption{Values of hyperparameters for different RL algorithms.}
\centering
\begin{tabular}{l|llllllll}
\toprule
Model & $\alpha_{\mathrm{critic}}$ & $\alpha_{\mathrm{actor}}$ & Optimiser & Batch size & $\gamma$ & bp & Memory size &  $\tau$ \\
\midrule
DQN   & 0.0001               & -                   & Adam      & 64         & 0.3             & 0.0020         & 5000        & 1000                  \\
PG    & -                    & 0.0001              & Adam      & -          & 0.3             & 0.0020         & -           & -                     \\
A2C   & 0.001                & 0.0001              & Adam      & 128        & 0.3             & 0.0020         & -           & -                 \\
\bottomrule
\end{tabular}
\label{table:parameter}
\end{table}

%LSTM effectively solves the gradient vanishing and exploding problem of recurrent neural networks by having cell struectures that keep a compact summary of past information, controlling memory retention with the forget gate and incorporating new information via the input gate. As such, the LSTM is able to learn representations of long-term relationships relevant to the tak -- sequentially updating its internal memory states with new observations at each step.

\subsection{Experimental Results}

We test both baseline models and our methods between 2011 and 2019, and we calculate the trade returns net of transaction costs as in Equation~\ref{eq:add_profit} for each contract. We then form a simple portfolio by giving equal weights to each contract, and the trade return of a portfolio is:
\begin{equation}
R_{t}^{\mathrm{port}} = \frac{1}{N} \sum_{i=1}^N R_t^i
\end{equation}
where $N$ represents the number of contracts considered and $R_t^i$ is the trade return for contract $i$ at time $t$. We evaluate the performance of this portfolio using following metrics as suggested in \cite{lim2019enhancing}:
\begin{enumerate}
\item $E(R)$: annualised expected trade return,
\item $\mathrm{std}(R)$: annualised standard deviation of trade return,
\item Downside Deviation (DD): annualised standard deviation of trade returns that are negative, also known as downside risk,
\item Sharpe: annualised Sharpe Ratio ($E(R) / \mathrm{std}(R)$),
\item Sortino: a variant of Sharpe Ratio that uses downside deviation as risk measures ($E(R) / \text{Downside Deviation}$),
\item MDD: maximum drawdown shows the maximum observed loss from any peak of a portfolio,
\item Calmar: the Calmar ratio compares the expected annual rate of return with maximum drawdown. In general, the higher the ratio is, the better the performance of the portfolio is, 
\item \% +ve Returns: percentage of positive trade returns,
\item $\frac{\text{Ave. P}}{\text{Ave. L}}$: the ratio between positive and negative trade returns.
\end{enumerate}

\begin{figure}[htb]
\begin{subfigure}{0.33\textwidth}
\includegraphics[width=\linewidth, height=3.6cm]{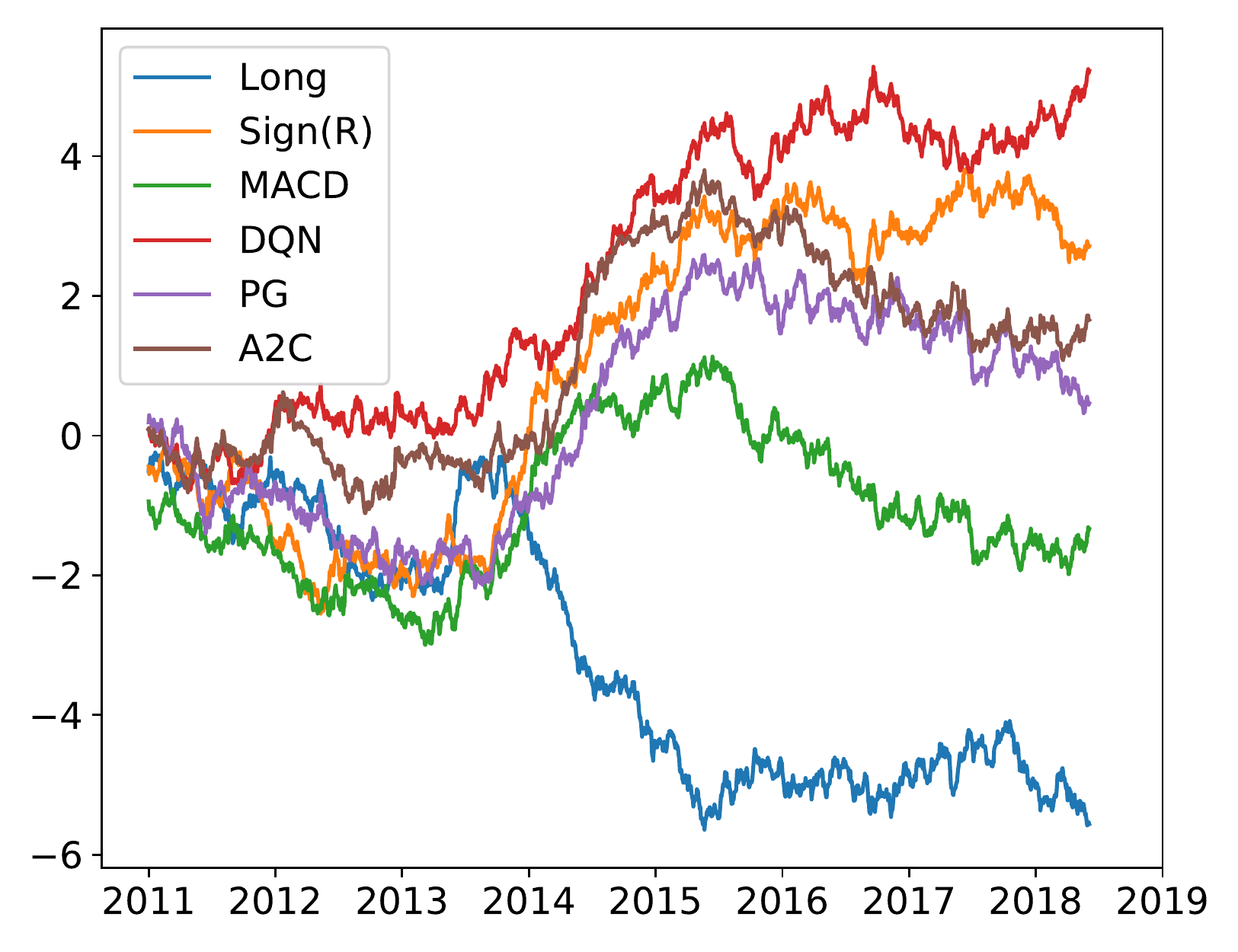} 
%\caption{Candlestick for daily data.}
%\label{fig:subim1}
\end{subfigure}
\begin{subfigure}{0.33\textwidth}
\includegraphics[width=\linewidth, height=3.6cm]{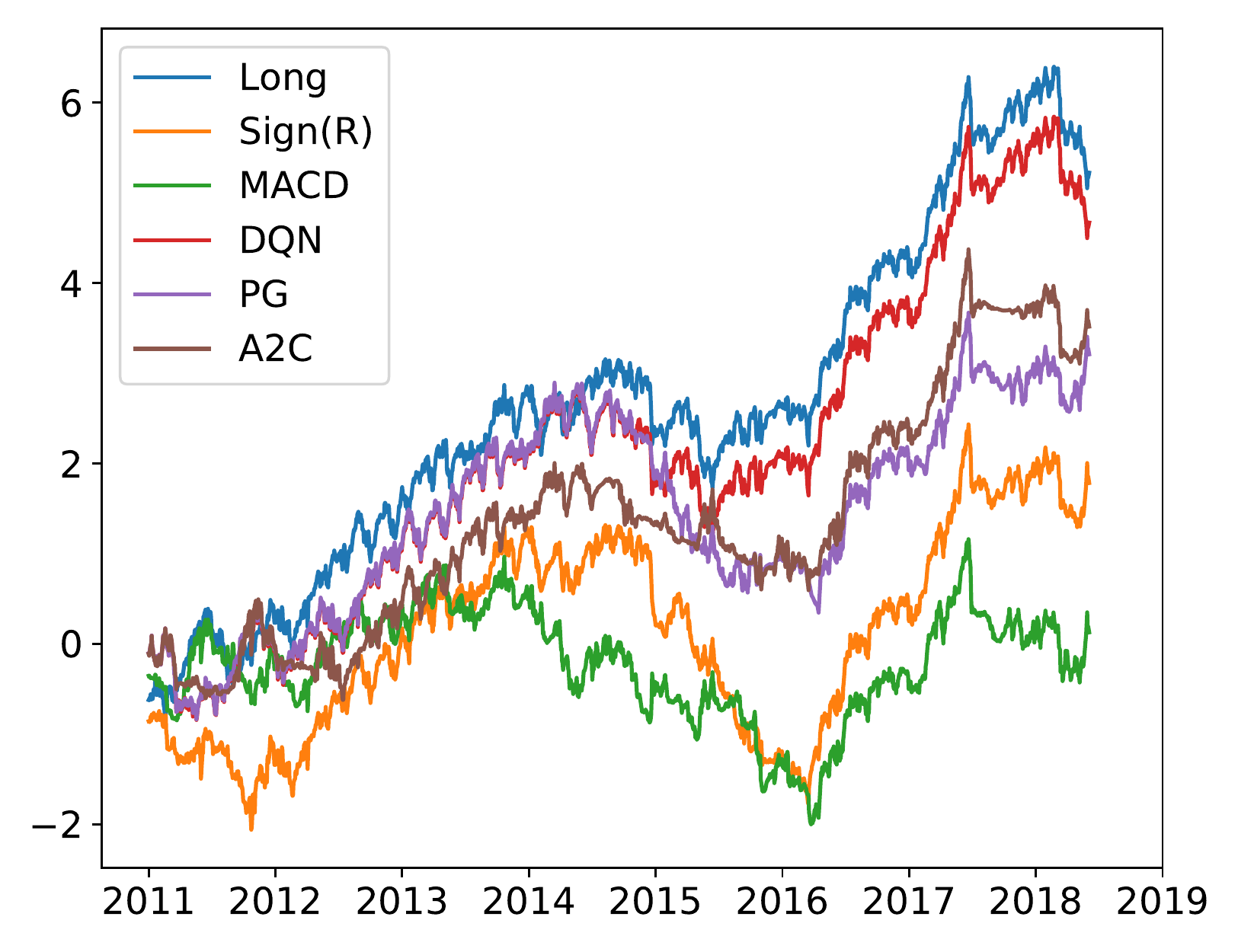}
\end{subfigure}
\begin{subfigure}{0.33\textwidth}
\includegraphics[width=\linewidth, height=3.6cm]{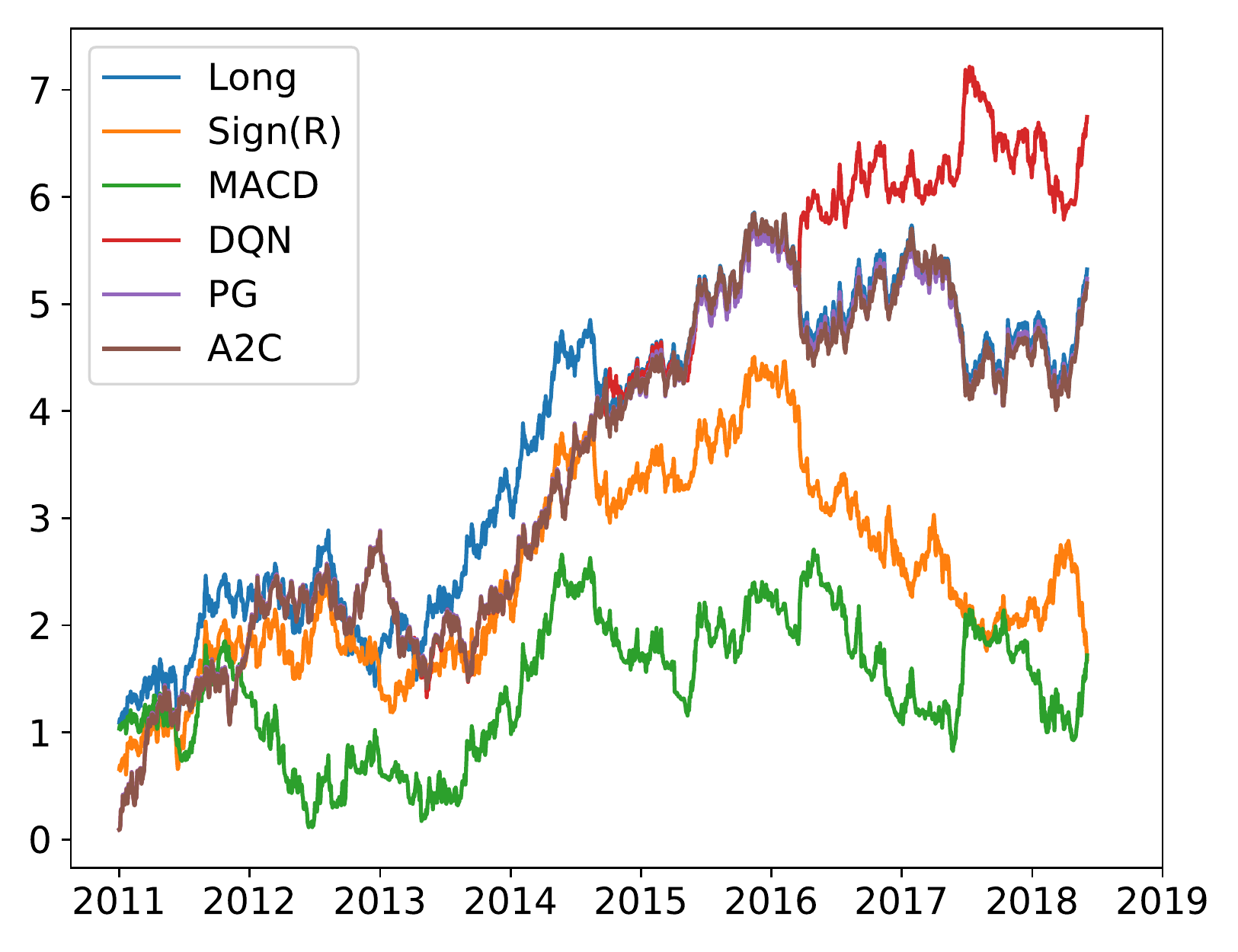}
\end{subfigure}

\begin{subfigure}{0.33\textwidth}
\includegraphics[width=\linewidth, height=3.6cm]{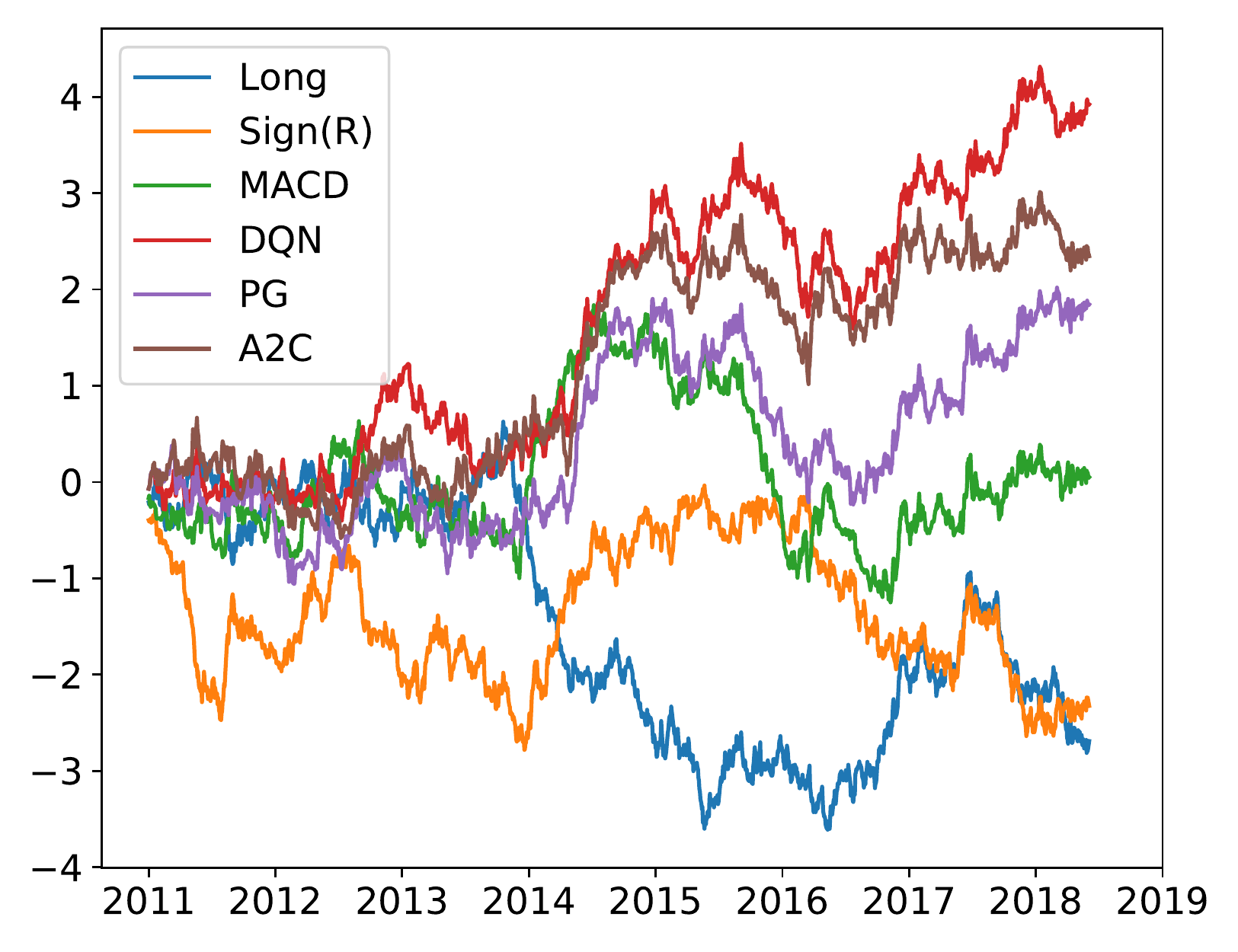}
\end{subfigure}
\begin{subfigure}{0.33\textwidth}
\includegraphics[width=\linewidth, height=3.6cm]{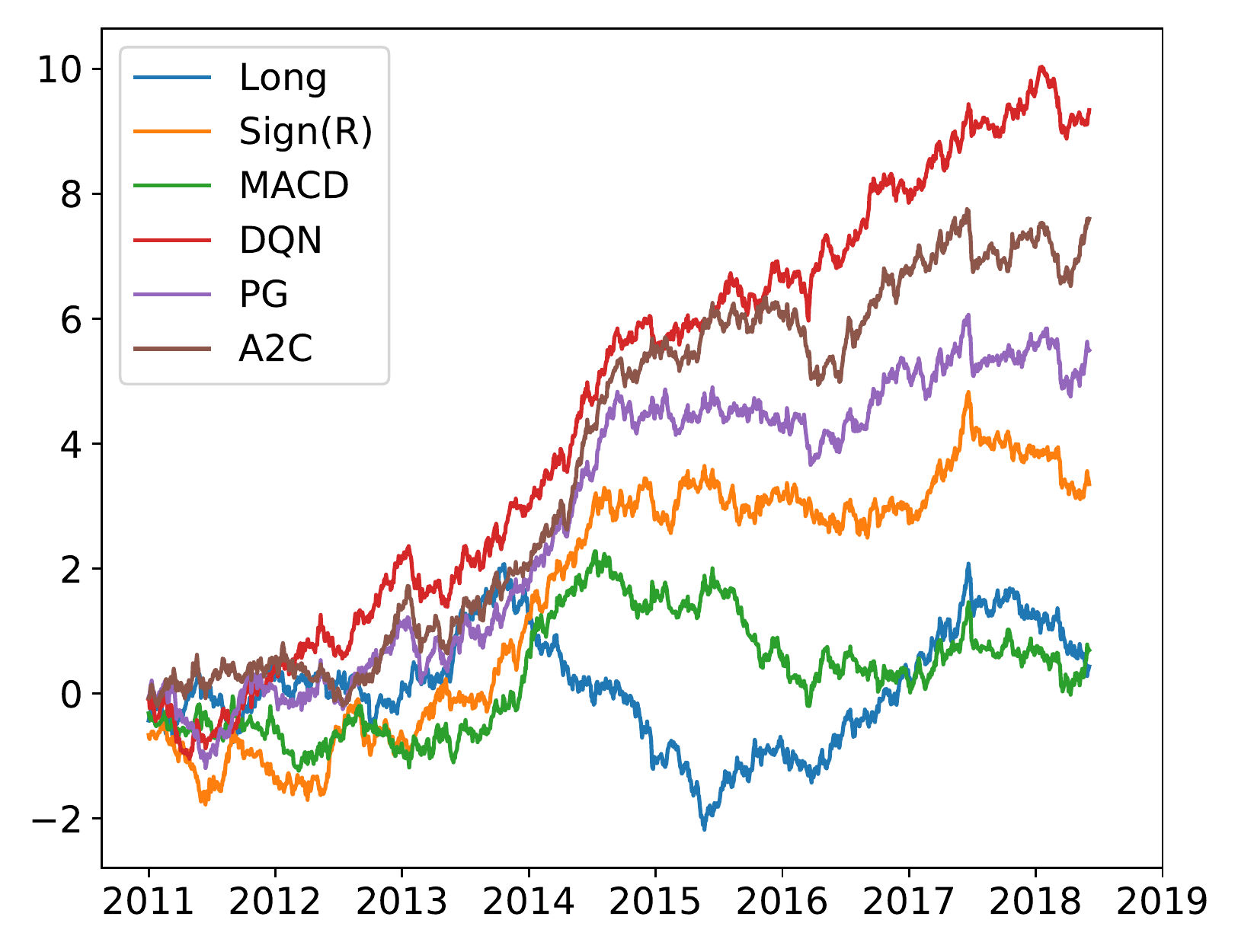}
\end{subfigure}
\caption{Cumulative trade returns for \textbf{First row:} commodity, equity index and fixed income; \textbf{Second row:} FX and the portfolio of using all contracts.}
\label{fig:cum_ret}
\end{figure}

We present our results in Table~\ref{table:result_target} where an additional layer of portfolio-level volatility scaling is applied for each model. This brings the volatility of different methods to a same target so we can direcly compare metrics such as expected and cumulative trade returns. We also include the results without this volatility scaling for reference in Table~\ref{table:result} in Appendix B. Table~\ref{table:result_target} is split into five parts based on different asset classes. The results show the performance of a portfolio by only using contracts from that spefic asset class. An exception is the last part of the table where we form a portfolio using all contracts from our dataset. The cumulative trade returns for different models and asset classes are presented in Figure~\ref{fig:cum_ret}.

\begin{figure}[htb]
\centering
\includegraphics[width=5.5in, height=2.2in]{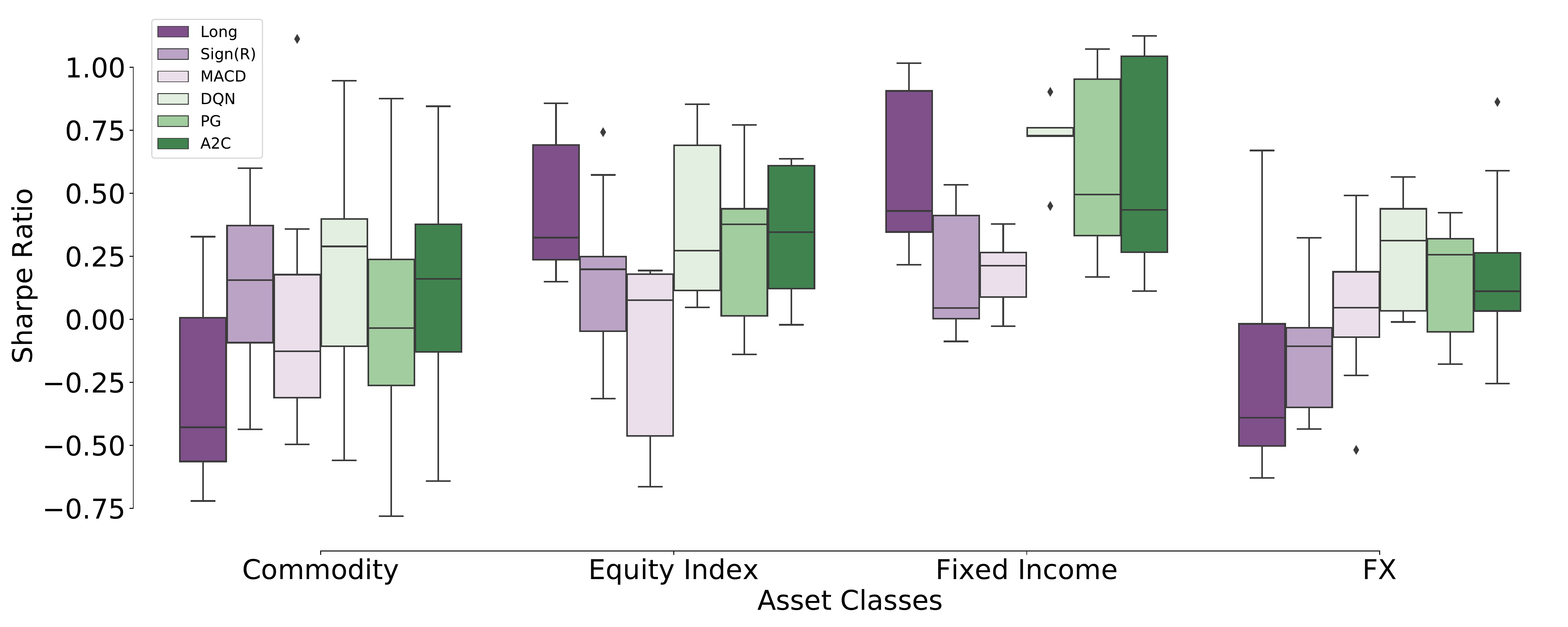}
\includegraphics[width=5.5in, height=2.2in]{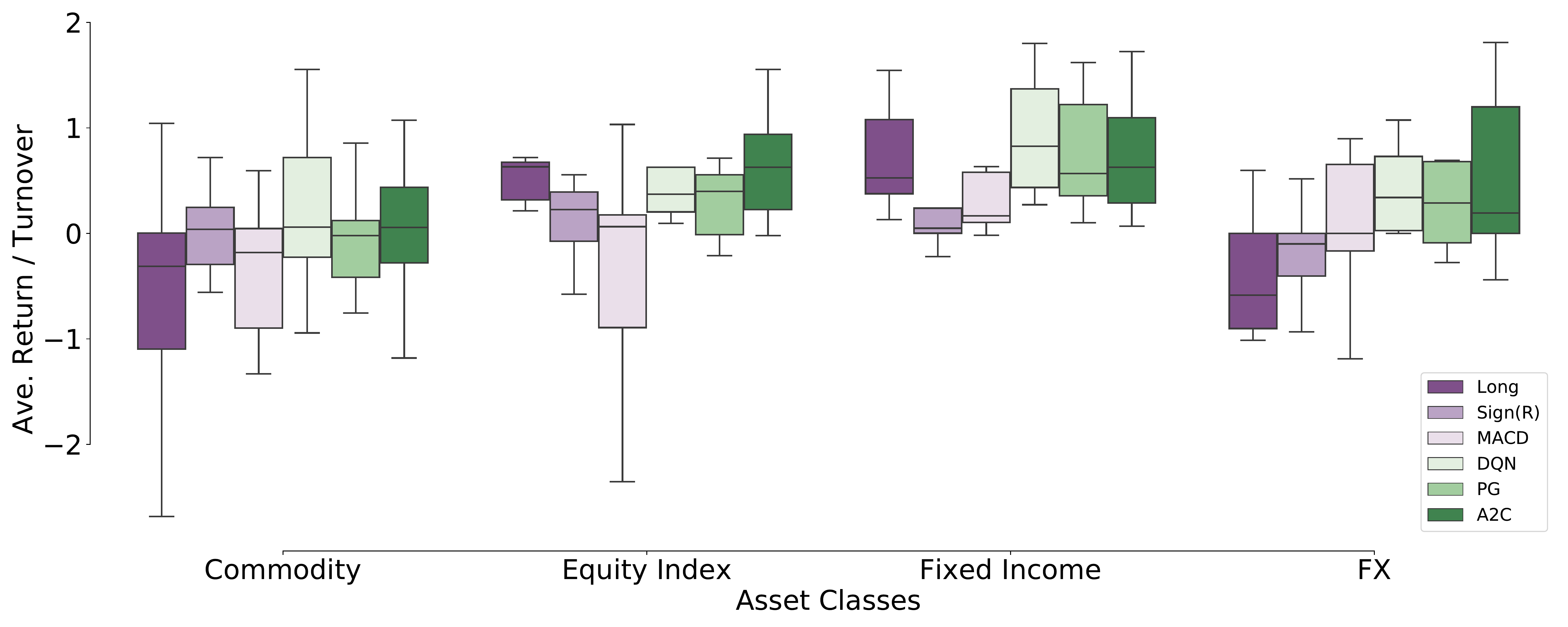}
\caption{Sharpe Ratio (\textbf{Top}) and average trade return per turnover (\textbf{Bottom}) for individual contracts.}
\label{fig:ind_sharpe}
\end{figure}

We can see that RL algorithms deliver better performance for most asset classes except for the equity index where a long-only strategy is better. This can be explained by the fact that most equity indices were dominated by large upward trends in the testing period. Similarly, fixed incomes had a growing trend until 2016 and entered into consolidation periods until now. Arguably, the most reasonable strategy we should have in these cases might be just to hold our positions if large trends persist. However, the long-only strategy performs the worst in commodity and FX markets where prices are more volatile. RL algorithms perform better in these markets being able to go long or short at reasonable time points. Overall, DQN obtains the best performance among all models and the second best is the A2C approach. We investigate the cause of this observation and find that A2C generates larger turnovers, leading to smaller average returns per turnover as shown in Figure~\ref{fig:ind_sharpe}. 

%%%%%%%%%%%%%%%%%%%%%%%%%%%%%%%%%%%%%%%%%%%%
%%%%%%%%%%%%%%%%%%%%%%%%%%%%%%%%%%%%%%%%%%%%
\begin{table}[t]
\caption{Experiment results for the portfolio-level volatility targeting.}
\label{table:result_target}
\centering
\begin{tabular}{l|lllllllll}
\toprule
              & \textbf{E(R)} & \textbf{Std(R)}   &  \textbf{DD} & \textbf{Sharpe} & \textbf{Sortino} & \textbf{MDD}   & \textbf{Calmar} & \textbf{\makecell{\% of\\+ Ret}} &  \textbf{$\frac{\text{Ave. P}}{\text{Ave. L}}$} \\
\midrule              
              & \multicolumn{9}{c}{\textbf{Commdity}}                                                                                    \\
\midrule              
Long      & -0.710   & 0.979  & 0.604   & -0.726  & -1.177  & 0.350  & -0.140   &0.473   & 0.989    \\
Sign(R) & 0.347    & 0.980   & 0.572   &0.354  & 0.606    & 0.116  &0.119     & 0.494   & 1.084  \\
MACD  &-0.171    & 0.978   & 0.584   & -0.175 & -0.293  & 0.190   &-0.060   & 0.486   &1.026     \\
DQN     &\textbf{0.703}     &0.973    &\textbf{0.552}    &\textbf{0.723}   & \textbf{1.275}     & \textbf{0.066}    & \textbf{0.501}   &  \textbf{0.498}   &  \textbf{1.135}  \\
PG         &0.062     &0.982     &0.585   &0.063   & 0.106    &0.039  &0.023   &0.495    & 1.029    \\
A2C      &0.223    &0.955   &0.559    &0.234  &0.399     &0.141   &0.091  &0.487    &1.093      \\
\midrule              
              & \multicolumn{9}{c}{\textbf{Equity Index}}                                                                                    \\
\midrule              
Long      & \textbf{0.668}    & 0.970   & 0.606   & \textbf{0.688}   & \textbf{1.102}   &0.132  & \textbf{0.509}     &\textbf{0.542}     & 0.948 \\
Sign(R) & 0.228     &  0.966  & 0.610  & 0.236   & 0.374   & 0.344  &0.077    &0.528   & 0.930        \\
MACD  & 0.016   & 0.962  & 0.618     &0.017  &0.027  & 0.311   & 0.006  & 0.519    &0.927    \\
DQN     & 0.629   & 0.970   &  0.606  & 0.648  & 1.038   & 0.161  &0.381   & 0.541    &0.944      \\
PG      & 0.432  & 0.967      &  0.605   & 0.447   &0.714    &0.242  & 0.185   &0.529   &  0.960   \\
A2C  & 0.473  & 0.929    & \textbf{0.593}    &0.510      & 0.798   &\textbf{0.124}  &0.328   &0.533    &\textbf{0.962}              \\
\midrule              
              & \multicolumn{9}{c}{\textbf{Fixed Income}}                                                                                    \\
\midrule              
Long        & 0.680   & 0.975   &  0.576    &0.698   & 1.180    & \textbf{0.061}    & 0.444    & 0.515   & 1.054   \\
Sign(R)  & 0.214  &  0.972   & 0.592    & 0.221    & 0.363   &  0.080 & 0.083    &  0.504  & 1.019   \\
MACD   & 0.219  & 0.967   & 0.579     &0.228   & 0.380   & 0.065   & 0.123    & 0.486   & \textbf{1.101}     \\
DQN      & \textbf{0.908}  &0.972   & \textbf{0.562}    &\textbf{0.935}     & \textbf{1.617}    & 0.062   &\textbf{0.543}  &0.515   &1.098     \\
PG          & 0.705  &0.974   &   0.576   & 0.724    & 1.225    &\textbf{0.061}     &0.436   &  \textbf{0.517}   & 1.052 \\
A2C       & 0.699  &0.979   &0.582 &0.714   & 1.203   &0.067   & 0.408 &\textbf{0.517}    & 1.048       \\
\midrule              
              & \multicolumn{9}{c}{\textbf{FX}}                                                                                    \\
\midrule              
Long       & -0.344   &  0.973  &  0.583  & -0.353  &  -0.590 &  0.423   & -0.097  &  0.491  & 0.979  \\
Sign(R) & -0.297   &0.973   &  0.592   &  -0.306  &-0.502    & 0.434    & -0.111    & 0.499   & 0.954  \\
MACD  & 0.006     &0.970   & 0.582    & 0.007    & 0.011   & 0.329   & 0.002     & 0.493   & 1.029    \\
DQN     &  \textbf{0.528}     & 0.967    &\textbf{0.553}  & \textbf{0.546} &\textbf{0.955}   &0.183  & \textbf{0.313}   &\textbf{0.510}    & \textbf{1.051}       \\
PG         &  0.248     & 0.967    &0.566   &0.257   &0.438  & 0.240  &0.124   &0.506   & 1.021     \\
A2C      &  0.316  & 0.963    & 0.563   &0.328    &0.561  &\textbf{0.165}  &0.201    &0.507  & 1.026              \\
\midrule              
              & \multicolumn{9}{c}{\textbf{All}}                                                                                    \\
\midrule              
Long      & 0.055    &   0.975    & 0.598  &  0.058    & 0.092  & 0.071  &  0.013   & 0.520  & 0.933 \\
Sign(R) & 0.429      &  0.972     & 0.582   & 0.441    &  0.737  &0.038  &0.201   & 0.510    &  1.031  \\
MACD  &  0.089    & 0.978    & 0.582    & 0.091   & 0.153   &0.008  & 0.035  &0.493  &  1.043     \\
DQN     &  \textbf{1.258}    &0.976    &\textbf{0.567}    & \textbf{1.288     } &\textbf{2.220 }   &\textbf{ 0.002}    &\textbf{1.025}      & \textbf{0.543} & \textbf{1.043}   \\
PG         & 0.740   &0.980     &0.593   &0.754    &1.247   &0.012   & 0.480      &  0.533   &0.990 \\
A2C      &1.024    & 0.975    &0.573   & 1.050    & 1.785   &0.007  & 0.685   & 0.538  &  1.021     \\
\bottomrule
\end{tabular}
\end{table}

\begin{figure}[H]
\centering
\includegraphics[width=5.5in, height=2in]{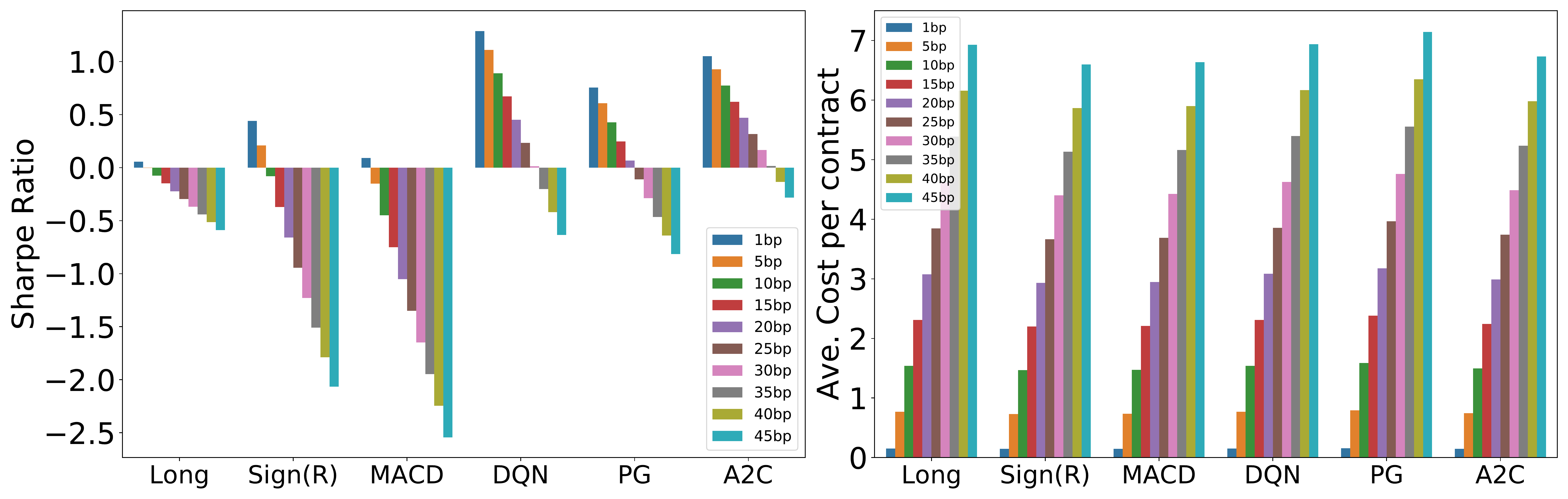}
\caption{Sharp Ratio (\textbf{Left} ) and average cost per contract (\textbf{Right}) under different cost rates.}
\label{fig:different_bp}
\end{figure}

We also investigate the performance of our methods under different transaction costs. In the left of Figure~\ref{fig:different_bp}, we plot the annualised Sharpe Ratio for the portfolio using all contracts at different cost rates. We can see that RL algorithms can tolerate larger cost rates and, in particular, DQN and A2C can still generate positive profits with cost rate at 25bp. To understand how cost rates (bp) translate into monetary values, we plot the average cost per contract at the right of Figure~\ref{fig:different_bp} and we can see that 25bp represents roughly \$3.5 per contract. This is a realistic cost for a retail trader to pay but institutional traders have a different fee structure based on trading volumes and often have cheaper cost. In any case, this shows the validity of our methods in a realistic setup. 

The performance of a portfolio is generally better than the performance of an individual contract as risks are diversified across a range of assets, so the return per risk is higher. In order to investigate the raw quality of our methods, we investigate the performance of individual contracts. We use the boxplots in Figure~\ref{fig:ind_sharpe} to present the annualised Sharpe Ratio and average trade return per turnover for each futures contract. Overall, these results reinforce our previous findings that RL algorithms generally work better, and the performance of our method is not driven by a single contract that shows superior performance, reassuring the consistency of our model.  

\section{Conclusion}
\label{conclusion}

We adopt RL algorithms to learn trading strategies for continuous futures contracts. We discuss the connection between modern portfolio theory and the RL reward hypothesis, and show that they are equivalent if a linear utility function is used. Our analysis focuses three RL algorithms, namely Deep Q-learning Network, Policy Gradients and Advantage Actor-Critic, and investigate both discrete and continuous action spaces. We utilise features from time series momentum and technical indicators to form state representations. In addition, volatility scaling is introduced to improve reward functions. We test our methods on 50 liquid futures contracts from 2011 to 2019, and our results show that RL algorithms outperform baseline models and deliver profits even under heavy transaction costs. 

In subsequent continuations of this work, we would like to investigate different forms of utility functions. In practice, an investor is often risk-averse and the objective is to maximise a risk-adjusted performance function such as the Sharpe Ratio, leading to a concave utility function. As suggested in \cite{huang2018financial}, we can resort to distributional reinforcement learning \cite{bellemare2017distributional} to obtain the entire distribution over $Q(s,a)$ instead of the expected Q-value. Once the distribution is learned, we can choose actions with the highest expected Q-value and the lowest standard deviation of it, maximising the Sharpe Ratio. We can also extend our methods to portfolio optimisation by modifying the action spaces to give weights of individual contracts in a portfolio. We can incorporate the reward functions with mean-variance portfolio theory \cite{markowitz1952portfolio} to deliver a reasonable expected trade return with minimal volatility.

\section*{Acknowledgement}
The authors would like to thank Bryan Lim, Vu Nguyen, Anthony Ledford and members of Machine Learning Research Group at the University of Oxford for their helpful comments. We are most grateful to the Oxford-Man Institute of Quantitative Finance, who provided the Pinnacle dataset and computing facilities.

%\bibliographystyle{plain}
%\bibliography{mybibliography.bib}

\newpage
\appendix
\section*{Appendix A}

Our dataset consists of 50 futures contracts, and there are 25 commodity contracts, 11 equity index contracts, 5 fixed income contracts and 9 forex contracts. A detail description of each contract is below:

\subsection{Commodities}

\begin{table}[htb]
\centering
\begin{tabular}{l|l}
\toprule
Ticker & Contract details          \\
\midrule
CC     & COCOA                     \\
DA     & .MILK III, Comp           \\
GI     & GOLDMAN SAKS C. I.        \\
JO     & ORANGE JUICE              \\
KC     & COFFEE                    \\
KW     & WHEAT, KC                 \\
LB     & LUMBER                    \\
NR     & ROUGH RICE                \\
SB     & SUGAR \#11                \\
ZA     & PALLADIUM, Electronic     \\
ZC     & CORN, Electronic          \\
ZF     & FEEDER CATTLE, Electronic \\
ZG     & GOLD, Electronic          \\
ZH     & HEATING OIL, Electronic   \\
ZI     & SILVER, Electronic        \\
ZK     & COPPER, Electronic        \\
ZL     & SOYBEAN OIL, Electronic   \\
ZN     & NATURAL GAS, Electronic   \\
ZO     & OATS, Electronic          \\
ZP     & PLATINUM, electronic      \\
ZR     & ROUGH RICE, Electronic    \\
ZT     & LIVE CATTLE, Electronic   \\
ZU     & CRUDE OIL, Electronic     \\
ZW     & WHEAT, Electronic         \\
ZZ     & LEAN HOGS, Electronic    \\
\bottomrule
\end{tabular}
\end{table}

\subsection{Equity Indexes}

\begin{table}[htb]
\centering
\begin{tabular}{l|l}
\toprule
Ticker & Contract details           \\
\midrule
CA     & CAC40 INDEX                \\
EN     & NASDAQ, MINI               \\
ER     & RUSSELL 2000, MINI         \\
ES     & S \& P 500, MINI           \\
LX     & FTSE 100 INDEX             \\
MD     & S\&P 400 (Mini Electronic) \\
SC     & S \& P 500, Composite      \\
SP     & S \& P 500, Day Session    \\
XU     & DOW JONES EUROSTOXX50      \\
XX     & DOW JONES STOXX 50         \\
YM     & Mini Dow Jones (\$5.00)    \\
\bottomrule
\end{tabular}
\end{table}

\subsection{Fixed Incomes}
\begin{table}[H]
\centering
\begin{tabular}{l|l}
\toprule
Ticker & Contract details           \\
\midrule
DT     & EURO BOND (BUND)          \\
FB     & T-NOTE, 5-year Composite  \\
TY     & T-NOTE, 10-year Composite \\
UB     & EURO BOBL                 \\
US     & T-BONDS, Composite      \\
\bottomrule 
\end{tabular}
\end{table}

\subsection{Forex}
\begin{table}[H]
\centering
\begin{tabular}{l|l}
\toprule
Ticker & Contract details           \\
\midrule
AN     & AUSTRALIAN, Day Session \\
BN     & BRITISH POUND, Composite   \\
CN     & CANADIAN, Composite     \\
DX     & US DOLLAR INDEX            \\
FN     & EURO, Composite            \\
JN     & JAPANESE YEN, Composite    \\
MP     & MEXICAN PESO               \\
NK     & NIKKEI INDEX               \\
SN     & SWISS FRANC, Composite    \\
\bottomrule
\end{tabular}
\end{table}

\newpage
\section*{Appendix B}

Table~\ref{table:result} presents the performance metrics for portfolios without additional layer of volatility scaling.

\begin{table}[htb]
\caption{Experiment Results for the Raw Signal.}
\label{table:result}
\centering
\begin{tabular}{l|lllllllll}
\toprule
              & \textbf{E(R)} & \textbf{Std(R)}   &  \textbf{DD} & \textbf{Sharpe} & \textbf{Sortino} & \textbf{MDD}   & \textbf{Calmar} & \textbf{\makecell{\% of\\+ Ret}} &  \textbf{$\frac{\text{Ave. P}}{\text{Ave. L}}$} \\
\midrule              
              & \multicolumn{9}{c}{\textbf{Commdity}}                                                                                    \\
\midrule              
Long      & -0.298   & 0.412  & 0.258   & -0.723  & -1.152   & 0.248  & -0.130   &0.473   & 0.987    \\
Sign(R) & 0.101      & 0.312   & 0.185   & 0.325  & 0.548     & 0.082  &0.115     & 0.494   & 1.081  \\
MACD  &-0.039    & 0.227   & 0.136   & -0.174 & -0.290   & 0.132   & -0.059   & 0.486   & 1.024     \\
DQN     &0.187     &0.301    &0.173    &0.623   & 1.085     & 0.039    &  0.413   &  0.498   &  1.119  \\
PG         &0.013     &0.287     & 0.172   & 0.047   & 0.078    & 0.072  &0.017   &0.495    & 1.026    \\
A2C      & 0.072    &0.163   &0.098    & 0.440  & 0.729     &0.099   & 0.161  &0.487    & 1.151      \\
\midrule              
              & \multicolumn{9}{c}{\textbf{Equity Indexes}}                                                                                    \\
\midrule              
Long      & 0.504    & 0.928   & 0.606   & 0.543   & 0.831   & 0.127  & 0.466     &0.541     & 0.928 \\
Sign(R) & 0.168     &  0.799  & 0.526    & 0.211   & 0.319   & 0.299    & 0.075    &0.528   & 0.928        \\
MACD  & -0.068   & 0.586  & 0.385    &-0.117  &-0.178  & 0.351     & -0.041  & 0.519    &0.904     \\
DQN     & 0.461     & 0.933   &  0.611    & 0.494  & 0.754   & 0.170   & 0.308   & 0.541    & 0.922      \\
PG      & 0.320  & 0.875      &  0.574    & 0.366   &0.558    &0.211   &  0.183   & 0.529   &  0.949   \\
A2C  & 0.293  & 0.629    & 0.427    &  0.466      & 0.686   & 0.193  & 0.214   &0.533    & 0.965              \\
\midrule              
              & \multicolumn{9}{c}{\textbf{Fixed Income}}                                                                                    \\
\midrule              
Long        & 0.605   & 0.939   &  0.561    &0.645   & 1.081    & 0.108    & 0.455    & 0.515   & 1.048   \\
Sign(R)  & 0.189  &  0.795   & 0.496    & 0.237    & 0.381   &  0.165 & 0.103    &  0.504  & 1.024   \\
MACD   & 0.136  & 0.609   & 0.367     & 0.224   & 0.371   & 0.124   & 0.131    & 0.485   & 1.102     \\
DQN      & 0.734  & 0.862   & 0.508    & 0.851     & 1.445    & 0.118   &   0.469  &0.515   &1.086     \\
PG          & 0.624  & 0.938   &  0.561   & 0.665    & 1.113    & 0.109     &0.443   &  0.517   &  1.043  \\
A2C       & 0.852  & 1.345   &  0.806  &  0.633   & 1.057   &0.128   & 0.397 &0.517    & 1.039       \\
\midrule              
              & \multicolumn{9}{c}{\textbf{FX}}                                                                                    \\
\midrule              
Long       & -0.198   &  0.472  &  0.285      & -0.420  &  -0.696 &  0.219   &  -0.101  &  0.491  &  0.966  \\
Sign(R) & -0.113      & 0.551   &  0.341   &  -0.207  &-0.332    & 0.170    & -0.071    & 0.499   & 0.968  \\
MACD  & 0.016      & 0.424   & 0.259    & 0.037    & 0.061   &  0.156   & 0.016     & 0.493   & 1.034    \\
DQN     &  0.272     & 0.487    &  0.280   & 0.560  &0.972   & 0.085  &  0.326   &0.510    & 1.058       \\
PG         &  0.157     & 0.533    &   0.312   & 0.295   & 0.503  & 0.098  & 0.148   &0.506   & 1.029     \\
A2C      &  0.159   & 0.455    & 0.267   & 0.349    &  0.592  & 0.081  & 0.193    &0.507  & 1.034              \\
\midrule              
              & \multicolumn{9}{c}{\textbf{All}}                                                                                    \\
\midrule              
Long      & -0.013    &   0.363    & 0.230  &  -0.036    & -0.057  & 0.037  &  -0.009   & 0.519  & 0.919 \\
Sign(R) & 0.086      &  0.296     & 0.186   & 0.291       &  0.461    &  0.016  & 0.142   & 0.510    &  1.008  \\
MACD  &  -0.018     & 0.230    & 0.143    & -0.080    & -0.129   & 0.026  &  -0.029  &0.493  &  1.013     \\
DQN     &  0.318     &0.252    &0.150    & 1.258      &2.111     &   0.008    &  1.023      & 0.543 & 1.041   \\
PG         & 0.168      &0.279     & 0.174   & 0.602    &  0.968   & 0.011   &   0.373      &  0.533   &  0.968 \\
A2C      & 0.214    &  0.221    &  0.134   & 0.969    &  1.601   & 0.009  & 0.672   & 0.538  &   1.014     \\
\bottomrule
\end{tabular}
\end{table}
%%%%%%%%%%%%%%%%%%%%%%%%%%%%%%

%\subsubsection*{Acknowledgments}

%Use unnumbered third level headings for the acknowledgments. All acknowledgments go at the end of the paper. Do not include acknowledgments in the anonymized submission, only in the final paper.

\end{document}